\documentclass[twocolumn,aps,prb,groupedaddress,showpacs]{revtex4}
\usepackage[OT1]{fontenc}
\usepackage{graphicx}

\def  \bsig    {\mbox{\boldmath$\sigma$}}
\begin{document}

\title{Ground state splitting of $^8S$ rare earth ions in semiconductors}

\author{A. {\L}usakowski}
\affiliation{Institute of Physics, Polish Academy of Sciences,
Al. Lotnik\'{o}w 32/46, 02-668 Warsaw, Poland}
\begin{abstract}
We propose a new mechanism leading to the ground state splitting for
the rare earth $^8S$ ions in semiconductor crystals. The resulting
splitting is due to three effects, the first is the intra atomic
$4f-5d$ spin-spin interaction, the second one is the spin - orbit
interaction for $5d$ electrons and the third one is their 
hybridization with the valence band states of semiconductors host. The
resulting splitting significantly depends on the relative 
position of $5d$ level with respect to semiconductor host band structure. We also discuss
different model, already known in the literature, which is  also based
on ion - band states hybridization. For both models, as an example, we
present results of numerical calculations for rare earth ion in IV-VI
semiconductor PbTe.
\end{abstract}
\maketitle
\section{Introduction}
\begin{figure}
\begin{center}
\includegraphics[scale=0.3,angle=0]{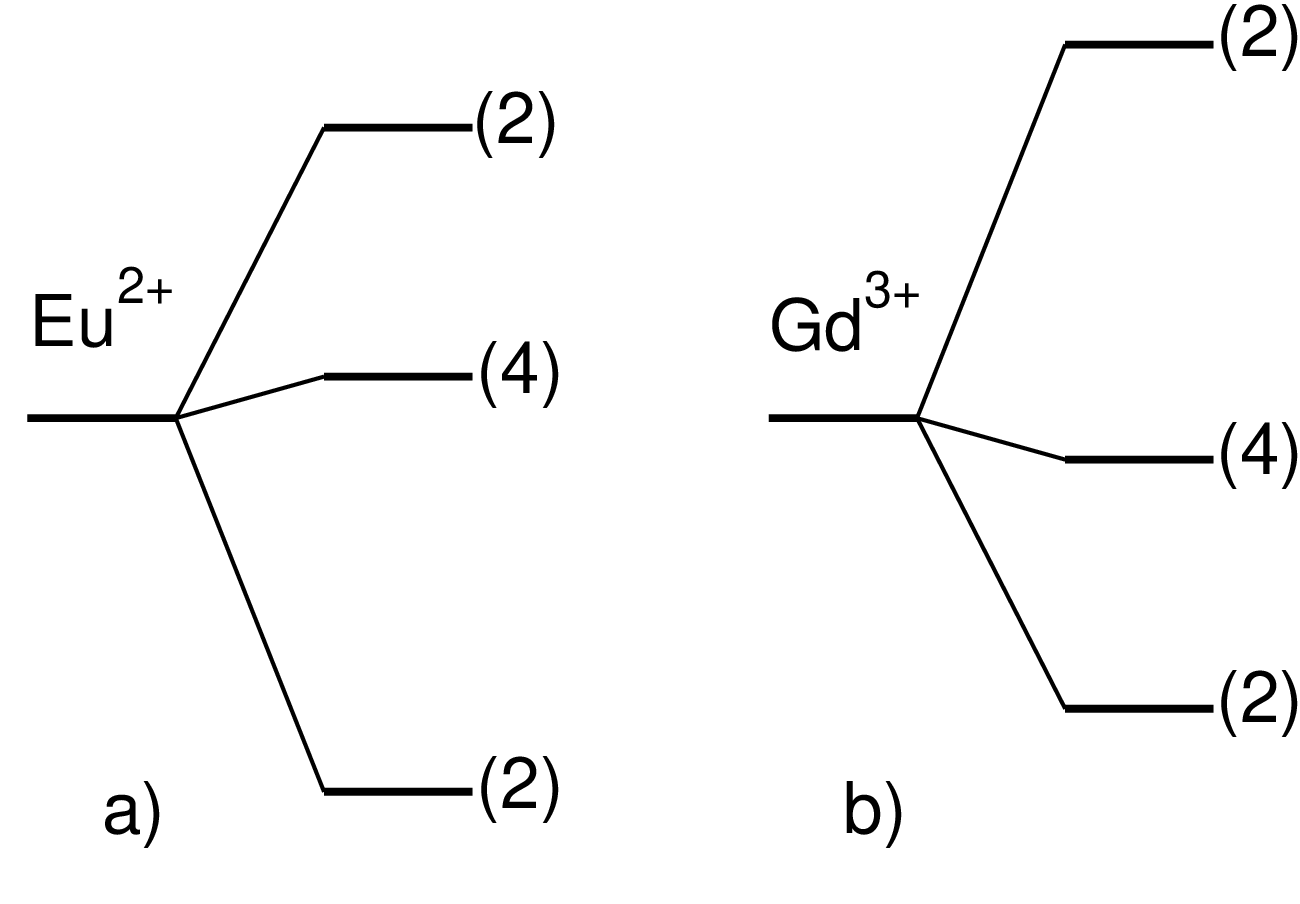}
\end{center}
\caption{\label{fig_spl} Differences in the splittings of the ground
state levels for a) Eu$^{2+}$ and b) Gd$^{3+}$ ions in PbTe
crystal. The numbers show the degeneracies of the levels. }
\end{figure}

The present paper is devoted to the theoretical analysis of the ground
state splitting of rare earth (RE) $^8S$ ions semiconductor
crystals. Very well known and studied examples of such ions
are  Eu$^{2+}$ and 
Gd$^{3+}$.  It is believed that the magnetic properties of these 
ions are determined by the half filled 4$f$ shell. According to the Hund's
rule, for such electron configuration, the ground state should be
characterized by the total angular momentum L=0 and the total spin 
S=7/2. In other words the ground state should be 
$^8$S$_{7/2}$ which is eight fold degenerate, independently of the
crystal environment. Such a model explains
quite well magnetic susceptibility or magnetization measurements where
the spin of the ion interacts with external magnetic field or with the
spin of another ion and the energies of 
Zeeman or ion-ion exchange interactions are of the order of 1 K. \\
However, the electron paramagnetic resonance (EPR) experiments, which
probe the system on a much finer energy scale, clearly
demonstrate that the above picture is an approximate one only. It
turns out that in crystals the degeneracy of the ion's ground state is lifted, and
the nature of the splitting depends on the symmetry of the
enviroment. For example, for $O_h$ symmetry, the case we deal in the
present paper with,  the ground
state is split into three levels, doublet, quartet and doublet. 
In accordance with the group theory, the
effective spin Hamiltonian describing the splitting caused by the ion's
crystal environment is of the form 
\begin{equation}
\label{intro1}
H=\frac{b_4}{60}\left(O^0_4+5O^4_4\right)+\frac{b_6}{1260}\left(O^0_6-21O^4_6\right)
\end{equation}
where the operators equivalents $O_k^m$ for spin S=7/2 are $8\times 8$
matrices 
defined, for example, in Ref. \onlinecite{abragam}. The energy levels of Hamiltonian
(\ref{intro1}) are $(-18b_4-12b_6)^{(2)}$, $(2b_4+16b_6)^{(4)}$,
$(14b_4-20b_6)^{(2)}$,
where the superscripts $^{(2)}$ or $^{(4)}$ denote the degeneracy of
the levels. 
\\
From EPR experiments, if the quality of the samples is high, it is
possible to obtain not only the absolute values of coefficients $b_4$
and $b_6$, but also their signs. This may be achieved by performing
measurements of the same sample in different temperatures. 
The splittings, together with levels' degeneracies  are 
shown schematically in Fig.~\ref{fig_spl} 
for $b_4>0$ and $b_4<0$. The energy diagrams in Fig.~\ref{fig_spl}
correspond to the splittings of  Eu$^{2+}$ and
Gd$^{3+}$ ions in PbTe. This compound  is an example we will often refer  in the
present paper to, but the calculational methods and the general
predictions of the theory may be applied to any semiconductor
containing RE $S$ state ions. From the literature we know, that for Eu
ion\cite{bacskay} $b_4$=129 MHz, while for Gd 
ion\cite{bartkowski} $b_4$=\ -110.16 MHz. The 
coefficient $b_6$ is usually about two orders of magnitude smaller and
because its influence on the ground state splitting is very small it will not be
analyzed in the following.  Notice that the signs of $b_4$ for
Eu$^{2+}$ and Gd$^{3+}$ ions are opposite. This is rather strange,
because at the first sight,  
except for the small relative difference of nuclear charges these two ions are
very similar and their crystal neighbourhoods are the same (six
tellurium atoms placed in the vertices of regular octahedron). \\
The problem of the ground state splitting of rare earth $^8S_{7/2}$
ions is very old. 
Very comprehensive discussion of the possible physical
mechanisms leading to the splitting is due to Wybourne.\cite{wybourne} In
Ref.~\onlinecite{wybourne}, using perturbation theory and results of
numerical calculations  he analyzed the
case of Gd$^{3+}$ ion in the crystal environment of $D_{3h}$
symmetry. The main idea is, that due to strong spin-orbit interaction
for $4f$ electrons the 
higher energy $4f^7$ states $^{2S+1}L_{7/2}$ with different $L$ and $S$ are
mixed into  the ion's ground state 
\begin{equation}
\label{intro2_a}
|^8S_{7/2}\rangle \rightarrow s|^8S_{7/2}\rangle+p|^6P_{7/2}\rangle
+d|^6D_{7/2}\rangle ...
\end{equation}
where $s$, $p$, $d$, ... are numerical coefficients.\cite{newman} 
Due to nonzero $p$, $d$, ...  the ion is no
longer in pure $^8S_{7/2}$ state, it may interact with the 
crystal field and this iteraction leads to the ground state splitting.\\
The results obtained by
Wybourne have been recently improved by Smentek~et.~
al.\cite{smentek}. On the basis of extensive numerical relativistic
calculations for Gd$^{3+}$ ion they derived the effective spin
Hamiltonian for this ion in the crystal neighborhood of arbitrary
symmetry (relativistic crystal field theory). In
particular, for $O_h$ symmetry the effective spin Hamiltonian may be
written in the form of Eq. (\ref{intro1}) where the coefficient $b_4$
reads 
\begin{equation}
\label{intro2}
b_4=\frac{A_0^4U^{(4)}}{6\sqrt{154}}\left(\frac{1}{\sqrt{2}}b_4(04)+b_4(13)X^{(13)4}
+b_4(15)X^{(15)4} \right)
\end{equation}
The definitions and numerical values of $U^{(4)}$, $b_4(\kappa t)$ and
$X^{(\kappa t)4}$ may be found in Ref.~\onlinecite{smentek}.
$A_0^4$ is the crystal field coefficient, which, in the simplest six point charge
model,\cite{sugano} is equal $A_0^4=\frac{7e^2a_0^4}{2d^5}$ where $a_0\approx$0.5
\AA \ is the atomic length unit, $d$ is the distance between cation and
anion and $e$ is the electron's charge. 
Taking Eu and Gd in PbTe as example, we see that the relativistic
crystal field theory alone cannot simultaneously 
explain ground state splitting for both ions because the
experimentally determined signs of $b_4$ are opposite. It is very
improbable that replacing Eu  
ion by Gd ion we change the sign of $A_0^4$ because the neighborhoods
of two ions are the same. We also do not expect
significant changes of the parameters $U^{(4)}$, $b_4(\kappa t)$ and
$X^{(\kappa t)4}$ because these two ions are very similar.  
\\
The inadequacy of a model in which the RE ion interacts with the
enviroment by electrostatic crystal field potential only has been
already noticed in the literature\cite{buckmaster,newman,barnes}. In
1978 Barnes et. al.\cite{barnes}, analysing the ground state splitting for
Gd$^{3+}$  ion in different crystals, noticed that in insulators there is proportionality
between the crystal field coefficient $A_0^4$ and coefficient $b_4$,
while in metals such proportionality does not exist. They proposed a
model in which $4f$ electrons interact with the band states via
hybridization processes. Using second order perturbation theory with
respect to the hybridization they
constructed effective spin Hamiltonian for the ground state of
Gd$^{3+}$ ion. As the excited states of the system they took into
account configurations in which the number of electrons on $4f$ shell
changes
by $\pm 1$, i. e. $4f^8$ configuration plus one hole in the Fermi sea 
or $4f^6$ configuration plus one additional electron in the
band. According to the Hund's rule, in the
excited states $4f^8$ and $4f^6$ the angular momentum is
nonzero. Taking into account internal spin - orbit coupling, the authors of
Ref.~\onlinecite{barnes}  obtained effective spin - lattice
interaction leading to the ground state splitting. \\
In Section IV we will rederive this model because we think that in
its derivation and solution presented in Ref.~\onlinecite{barnes} some important
points have been missed.
\\
The main idea of Barnes et. al. is very interesting, however the final
results strongly depend on the constants describing the hybridization
between $4f$ electrons and the band states. Due to strong
localization of $4f$ shell, in the literature, it is very often
assumed that it is very small. For example in calculations of the
exchange integral between  $4f$ spin and band carriers in PbEuTe,
Dietl at. al.\cite{dietl1} found that the contribution to the final result
resulting from the hybridization $4f$- band states processes is
negligible.  \\
Contrary to $4f$ electrons, $5d$ states of RE ions are very extended
in space and their hybridization with band states is certainly much
stronger. In the literature we have found  examples of successful
explanations of magnetic properties of RE ions in semiconductors which
are based on the assumption that the interaction between $4f$
electrons and band states goes via internal $4f-5d$ exchange
interaction and $5d$ - band states hybridization. For instance, using
such a model, in 1970 Kasuya\cite{kasuya} explained the Eu-Eu 
exchange constant in EuO. The Kasuya mechanism was
used also, as a starting point, by 
Story et.~al.\cite{story1} in the theory explaining the 
Fermi energy dependence of the Gd-Gd exchange constant in SnGdTe mixed
crystals. The 4$f$~-~5$d$ interaction was also invoked by Dietl
et.~al.\cite{dietl1} in calculation of $sp-f$ exchange integral between
localized Eu spin and the band carriers in PbEuTe. 
\\

In the present paper we generalize this mechanism by including
spin-orbit coupling and crystal field potential for $5d$ states and
apply it to the calculation of the 
coefficient $b_4$. In the model, which will be explained and discussed
in detail in the
next sections, the ground state splitting
occurs due to combined effect of the intra-atomic, Heisenberg type,
4$f$-5$d$ exchange 
interaction, spin-orbit interaction on 5$d$  orbitals and the
hybridization of RE 5$d$ levels  with the valence band states. Our
model takes automatically the spin-orbit interaction in the
semiconductor host band states. It
turns out that the functional dependence of $b_4$ on $5d$ spin-orbit
constant is different for semiconductors with strong band spin-orbit
interaction from those for which this interaction may be
neglected. For semiconductors with strong band spin-orbit effects one
may expect that the proposed mechanism is more effective.  One of
the important parameters of the model is 
$\epsilon_0$, the energy necessary to transfer an electron 
from the valence band to the 5$d$ shell. The
magnitude of the resulting 4$f$ ground state splitting   
decreases very quickly with $\epsilon_0$.  That is why the position of
$5d$ level with respect to the Fermi energy decides whether the
mechanism is important or not.  For example, as it will be discussed
in Section III, for Gd in PbTe this 
energy is small, of the order of 0.5 eV\cite{story2} and the resulting
$b_4$ is of the order of the one observed in experiment, while for Eu 
$\epsilon_0$ is a few times larger and the calculated splitting is much
smaller. \\
In the next Section we describe the model and we derive approximate,
analytical result for $b_4$. This formula enables us to discuss the
salient features of the model, in particular the dependencies of $b_4$
on the parameters of the theory. In Section III we present details of
numerical calculations and Section IV is devoted to rederivation and
discussion of the model by Barnes et. al.\cite{barnes}. Some
additional remarks are contained in the last Section. 

\section{The effective spin Hamiltonian}
Let us consider a semiconductor crystal with one cation replaced by RE
atom. The unperturbed part of our model Hamiltonian describes ground
and  excited states of the system. These two groups of states are
connected by $5d$ - band states hybridization, which we treat here as a
perturbation. Below, we describe these two parts of Hamiltonian. \\
In the ground state of the system the electrons fill the band levels up
to the Fermi energy. We assume that the band structure is not changed
significantly by the presence of RE atom. In our model RE atom is
treated as the host cation atom with additional $4f$ and $5d$
orbitals.  There are seven electrons on $4f$ orbitals and $5d$ shell
is empty. Assuming validity of the Hund's rule, the spin of the ion 
$S=7/2$ and its angular momentum $L=0$ thus  the
ground state of the system is eight fold degenerate and is described
by $M$, the projection of spin on a quantization axis. \\ 
As the excited states we take configurations with one additional
electron on the $5d$ level and one electron less in the band. The
Hamiltonian describing $4f^75d^1$ configuration of the ion reads:
\begin{equation}
\label{sec1_1}
H_{4f^75d^1}=-J{\bf S}\cdot {\bf s} + \lambda_t {\bf L}_t\cdot {\bf s} 
+ \lambda_e {\bf L}_e \cdot {\bf s}
+ V_{cr} 
\end{equation}
The first term describes the exchange interaction between the $4f$
spin ${\bf S}$ and the spin ${\bf s}=\frac{1}{2} \bsig$ of the $5d$
electron. The second and the 
third terms describe the spin orbit interaction on the $5d$ shell. 
The operators $L_{ti}$ ($i=x,y,z$) are the angular momentum operators between
the $5d$ states of $t_{2g}$ symmetry and $L_{ei}$ between the $t_{2g}$
and $e_{2g}$ states respectively. For the definition of $t_{2g}$
($d_{yz}$, $d_{xz}$, $d_{xy}$) and
$e_{g}$ ($d_{z^2}$, $d_{x^2-y^2}$) states according to
Ref.~\onlinecite{sugano} they have following form:
\begin{widetext}
\begin{eqnarray}
\label{sec1_2}
L_{tx}=\left[\begin{array}{ccccc}0&0&0&0&0\\
0&0&i&0&0\\0&-i&0&0&0\\0&0&0&0&0\\0&0&0&0&0\\\end{array}\right]
L_{ty}=\left[\begin{array}{ccccc}
0&0&-i&0&0\\0&0&0&0&0\\i&0&0&0&0\\0&0&0&0&0\\0&0&0&0&0\\\end{array}\right]
L_{tz}=\left[\begin{array}{ccccc}0&i&0&0&0\\
-i&0&0&0&0\\0&0&0&0&0\\0&0&0&0&0\\0&0&0&0&0\\\end{array}\right]
\\
L_{ex}=\left[\begin{array}{ccccc}0&0&0&-i\sqrt{3}&-i\\
0&0&0&0&0\\0&0&0&0&0\\i\sqrt{3}&0&0&0&0\\i&0&0&0&0\\\end{array}\right]
L_{ey}=\left[\begin{array}{ccccc}0&0&0&0&0\\
0&0&0&i\sqrt{3}&-i\\0&0&0&0&0\\0&-i\sqrt{3}&0&0&0\\0&i&0&0&0\\\end{array}\right]
L_{ez}=\left[\begin{array}{ccccc}0&0&0&0&0\\
0&0&0&0&0\\0&0&0&0&2i\\0&0&0&0&0\\0&0&-2i&0&0\\\end{array}\right]
\end{eqnarray}
\end{widetext}
In the case of free ion  two spin orbit constants $\lambda_t$ and
$\lambda_e$  are equal, however if the ion is placed into the crystal they are
in general different.\cite{sugano} Finally, the last term in the Eq.~(\ref{sec1_1})
describes the influence of the crystal field on the $5d$ energy
levels and it has the form
\begin{equation}
\label{sec1_3}
V_{cr}=Dq\left[\begin{array}{ccccc}-4&0&0&0&0\\
0&-4&0&0&0\\0&0&-4&0&0\\0&0&0&6&0\\0&0&0&0&6\end{array}\right]
\end{equation}
After diagonalizing of the 80$\times$80 matrix $H_{4f^75d^1}$ we
obtain eigenvectors $|R\rangle$ and the corresponding eigenvalues
$\epsilon_R$. The eigenvectors $|R\rangle$ may be expressed in the
basis $|Md_i\sigma\rangle$:
\begin{equation}
\label{sec1_4}
|R\rangle = \sum_{Md_i\sigma}|Md_i\sigma\rangle\langle Md_i\sigma|R\rangle
\end{equation}
where $-7/2 \le M \le 7/2$ and $\sigma = \pm \frac{1}{2}$ are the
projections of the $4f$ and $5d$ spins on the
quantization axis, respectively and $d_i$ are $t_{2g}$ for $i=1,2,3$ and
$e_g$ for $i=4,5$ $5d$ orbitals.\\
In our model we
assume that the $5d$ levels of RE  hybridize with the band
states. (The hybridization of the $4f$ shell in the present Section is
neglected.) The hybridization matrix elements $\langle
R,q |h| M\rangle$ describe the probability amplitude of a transition
from  ground state $|M\rangle$ to excited state $|R,q\rangle$ with $h$
being one-electron hybridization Hamiltonian and the quantum number
$q\equiv n{\bf k}$ describes the band  state of an electron with the
corresponding energy $\epsilon_q$ transferred to $5d$ shell. The wave
vector ${\bf k}$ belongs to the first Brillouin zone and $n$ is
additional index necessary to fully characterize the band state. For
bands with negligible spin orbit coupling, where the electron spin is
a good quantum number, $n$ corresponds to the band's number and the
projection of electron's spin. \\ 
In the second order perturbation theory with respect to the hybridization between $5d$
levels of RE ion and the band states, which often is called
Schrieffer-Wolff transformation, we obtain the effective spin 
Hamiltonian for the RE ion in the crystal:
\begin{equation}
\label{sec1_5}
H_{MM'}=-\sum_{R,q}\frac{\langle M|h|R,q\rangle \langle R,q |h| M'\rangle}
{\epsilon_0+\epsilon_{R}-\epsilon_{q}}
\end{equation}
In  Eq.~(\ref{sec1_5}) the sum over $q$ runs over all occupied band
states. The sum in the denominator is the energy of the excited
state of the system where $\epsilon_0$ denotes the energy necessary
to transfer an electron from the Fermi level, which we
assume to be zero of the energy scale, to the lowest energy state
of $4f^75d^1$ configuration.\\
Using Eq.~(\ref{sec1_4}) and the fact that 
\begin{equation}
\label{sec1_6}
\langle Md_i\sigma q|h|M'\rangle=\delta_{MM'}\langle d_i\sigma|h|q\rangle
\end{equation}
Eq~(\ref{sec1_5}) may rewritten in the form
\begin{widetext}
\begin{equation}
\label{sec1_7}
H_{MM'}=-\sum_{R,q}\sum_{d_{i1}\sigma_1,d_{i2}\sigma_2}\frac{\langle d_{i2}\sigma_2|h|q\rangle
\langle q|h|d_{i1}\sigma_1\rangle \langle Md_{i1}\sigma_1
|R\rangle
\langle R|M'd_{i2}\sigma_2 \rangle
}
{\epsilon_0+\epsilon_{R}-\epsilon_{q}}
\end{equation}
\end{widetext}
Eq.~(\ref{sec1_7}) is the main result of the present paper. If we know
the values of matrix elements $H_{MM'}$ then, comparing them to Eq.~(\ref{intro1}),
we  obtain the coefficient $b_4$. In the next section we
present details and results of numerical calculations for the example
case of PbTe semiconductor, here we 
derive approximate, but analytical formula for $b_4$ valid to the fourth
order with respect to the intra-atomic spin-orbit coupling. This analytical formula
will enable us to discuss and understand the general properties and dependencies of
$b_4$ on different parameters of the model. \\ 
Let us take an arbitrary state characterized by wave vector ${\bf k}_0$
from the first Brillouin zone 
and let us denote by $\{{\bf k}_0\}$ the set of states which may be obtained
from ${\bf k}_0$ by symmetry transformations of $O_h$, including Kramers
conjugation. More precisely, we take an arbitrary wave vector ${\bf k}_0$
from the first Brillouin zone, we find all the wave vectors which may
be obtained from ${\bf k}_0$ by symmetry operations and the set
$\{{\bf k}_0\}$ contains the Kramers conjugate pairs corresponding to
these vectors. On the symmetry ground, the matrix 
$ Z_{d_{i1}\sigma_1,d_{i2}\sigma_2}\equiv \sum_{q \in \{{\bf k}_0\}} \langle
d_{i1}\sigma_1|h|q\rangle \langle
 q|h|d_{i2} \sigma_2\rangle$, appearing in Eq.~(\ref{sec1_7}), must have following form:
\begin{eqnarray}
\label{sec1_8}
Z_{d_{i1}\sigma_1,d_{i2}\sigma_2}=p_tI_t+p_eI_e+q_t {\bf L}_t\cdot \bsig 
+ q_e {\bf L}_e \cdot \bsig
\end{eqnarray}
where the matrices $I_t$ and $I_e$ are unit operators in $t_{2g}$ and
$e_g$ subspaces, respectively, i. e. :
\begin{equation}
\label{sec1_9}
I_t=\left[\begin{array}{ccccc}1&0&0&0&0\\
0&1&0&0&0\\0&0&1&0&0\\0&0&0&0&0\\0&0&0&0&0\\\end{array}\right]
I_e=\left[\begin{array}{ccccc}0&0&0&0&0\\
0&0&0&0&0\\0&0&0&0&0\\0&0&0&1&0\\0&0&0&0&1\\\end{array}\right]
\end{equation}
The coefficients $p_t$,$p_e$,$q_t$,$q_e$ depend on the band
structure and on the quantum number ${\bf k}_0$. If we neglect the band
spin-orbit interaction than 
$q_t$ and $q_e$ disappear. However, in many cases, for example for PbTe, such an
assumption is unjustified and, as we will see below, $q_t$ and $q_e$, in some
sense, play the role of ion's spin-orbit constants. \\
Let us notice that $\epsilon_q=\epsilon_{{\bf k}_{0}}$=const for all
$q\in\{{\bf k}_0\}$. Than the effective Hamiltonian $H_{MM'}$, Eq.~(\ref{sec1_7}), may be
rewritten as the trace over $d_i$ and $\sigma$ degrees of freedom of
the product of two matrices:
\begin{equation}
\label{sec1_10}
H_{MM'}=\frac{1}{48}\sum_{{\bf k}_0} {\rm Tr}_{d_i\sigma} Z Q(M,M')
\end{equation}
where the matrix $Q(M,M')$ is defined as
\begin{equation}
\label{sec1_11}
Q_{d_{i1}\sigma_1,d_{i2}\sigma_2}(M,M')=-\sum_{R}\frac{\langle Md_{i1}\sigma_1
|R\rangle
\langle R|M'd_{i2}\sigma_2 \rangle
}
{\epsilon_0+\epsilon_{R}-\epsilon_{{\bf k}_0}}
\end{equation}
and the factor 1/48 is necessary to take into account multiple counting of states
(the $O_h$ group has 48 elements). \\
Eq.~(\ref{sec1_11}) may be expanded in the series of $\lambda_t$ and
$\lambda_e$. Denoting $H_{\lambda}=\lambda_t {\bf L}_t\cdot {\bf s} 
+ \lambda_e {\bf L}_e \cdot {\bf s}$ the expansion is according to the
formula: 
\begin{equation}
\label{sec1_12}
\frac{1}{E-\frac{1}{2}J {\bf S}\cdot \bsig
+V_{cr}+H_{\lambda}}=G_0\sum_{n=0}^{\infty}\left(-H_{\lambda}G_0\right)^n
\end{equation}
where $E=\epsilon_0-\epsilon_q +7/4J+4Dq$. Notice that the lowest
eigenvalue of the operator $-\frac{1}{2}J{\bf S}\cdot \bsig+V_{cr}$ equals
$-(7/4J+4Dq)$ and the subtraction  of this term is in accordance with
the definition of $\epsilon_0$. The operator $G_0\equiv
1/\left(E+\frac{1}{2}J {\bf S}\cdot 
\bsig+V_{cr}\right)$ is equal:
\begin{equation}
\label{sec1_13}
G_0=\left(A_t+B_t{\bf S}\cdot \bsig\right)I_t+
\left(A_e+B_e{\bf S}\cdot \bsig\right)I_e
\end{equation}
with 
\begin{eqnarray}
\label{sec1_14}
A_t=\frac{\epsilon_0-\epsilon_q+\frac{9}{4}J}{\left(\epsilon_0-\epsilon_q\right)
\left(\epsilon_0-\epsilon_q+4J\right)}
\nonumber \\
\ \\
B_t=\frac{J}{2\left(\epsilon_0-\epsilon_q\right)\left(\epsilon_0-\epsilon_q+4J\right)}
\nonumber \\
\ \nonumber \\
\ \nonumber \\
A_e=\frac{\epsilon_0-\epsilon_q+10Dq+\frac{9}{4}J}{\left(\epsilon_0-\epsilon_q+10Dq\right)\left(\epsilon_0-\epsilon_q+10Dq+4J\right)}
\nonumber\\
\ \\
B_e=\frac{J}{2\left(\epsilon_0-\epsilon_q+10Dq\right)\left(\epsilon_0-\epsilon_q+10Dq+4J\right)}\nonumber
\end{eqnarray}
Calculations of traces in Eq~(\ref{sec1_10}) is a simple but very
tedious task, which may be simplified by use of a computer
program. We collect terms proportional to 
\begin{equation}
\label{sec1_15}
S_x^4+S_y^4+S_z^4=\frac{1}{20}\left(O^0_4+5O^4_4\right)+\frac{2331}{16}
\end{equation}
and we obtain the coefficient $b_4$, up to the fourth
power in ion's spin-orbit coupling: 
\begin{eqnarray}
\label{sec1_16}
b_4=-\frac{3}{2}\sum_{{\bf k}_0}B_t^2B_e\lambda_e\left\{q_tB_t\lambda_t\lambda_e
+q_e\left(B_t\lambda_t^2-2B_e\lambda_e^2\right) \right . \nonumber \\
\left . 
-\frac{1}{2}
\lambda_e\left[2q_t\left(A_t-A_e\right)B_t\lambda_e^2+8q_e\left(A_eB_t-A_tB_e\right)\lambda_t\lambda_e
\right . \right . \nonumber \\
\left .\left . 
+p_tA_t\left(3B_t\lambda_t^2-2B_e\lambda_e^2\right)
+p_eA_e\left(B_t\lambda_t^2-2B_e\lambda_e^2\right)
\right] \right\} . \nonumber \\
\ 
\end{eqnarray}
Let us discuss the main features of the obtained formula.\\
First, let us notice that $b_4$=0 for $\lambda_e$=0. This is the
general result valid for all orders of perturbation theory. It is
connected with the fact that if $\lambda_e$=0 there is no spin orbit
coupling between $t_{2g}$ and $e_{g}$ orbitals and this coupling is
the only one which mixes these two groups of states in Hamiltonian
$H_{\lambda}$. The operator 
$G_0$, Eq~(\ref{sec1_13}) does not mix $t_{2g}$ and $e_{g}$ states
too and, consequently, the same holds for operator
$Q_{d_{i1}\sigma_1,d_{i2}\sigma_2}(M,M')$, see
Eqs.~(\ref{sec1_11},\ref{sec1_12}). Notice, that for $\lambda_e$=0, in
Eq.~(\ref{sec1_12}), in the subspace of $e_g$ states, only the zeroth
order term survives and this term will certainly not lead to terms
in the effective spin Hamiltonian which have $O_h$ symmetry. Moreover,
due to this decoupling, the trace in Eq.~(\ref{sec1_10}) does not
depend on $q_e$. Thus, concerning the orbital degrees of freedom, we
may limit the considerations to the $t_{2g}$ subspace. But in this
subspace the problem is completely symmetrical with respect to
operations of full rotational group because $L_{tx}$,$L_{ty}$,$L_{tz}$
satisfy the angular momentum commutation relations for L=1 and there
is no reason to obtain $O_h$ terms in the effective Hamiltonian. We conclude
that the nonzero spin-orbit coupling $\lambda_e$ connecting the
$t_{2g}$ and $e_{g}$ states is the most important parameter of the
model. \\
Secondly, notice that the order of ion's spin-orbit coupling, $\lambda$, in
which we obtain nonzero $b_4$ depends on the band spin-orbit
coupling. If the spin-orbit coupling is absent in
the band, $q_t=q_e$=0, the lowest order terms are proportional to
$\lambda^4$. For nonzero band spin-orbit coupling they are proportional
to $\lambda^3$. In this sense the band spin-orbit coupling plays the
role of ion's spin-orbit coupling.\\
The formulas for $A_t$, $B_t$, $A_e$, $B_e$ show that the coefficient $b_4$ very
quickly decays with the excitation energy $\epsilon_0$. On the other
hand, 
these formulas suggest also that the most important contribution to the
final results comes from the hybridization of $5d$ level with the
states from the vicinity of the Fermi level. \\
Finally let us notice that the crystal field potential for $Dq>0$
diminishes $b_4$. This may be understood on the basis of the
preceding discussion devoted to the role of $\lambda_e$. Enlarging
the energy distance between $t_{2g}$ and $e_{g}$ states leads to the
effective decreasing of the coupling between these two groups of
states. 
\section{Results}
In this Section we show the example numerical calculations for RE ion
in PbTe semiconductor. \\
The electron band wave functions and band energies $\epsilon_q$  are
calculated according to the tight binding 
model developed in Ref.~\onlinecite{bauer}. In this model the band
states are build from $p$ and $s$ orbitals of Pb and Te. 
For a given momentum {\bf
k} belonging to the first Brillouin zone, the tight binding
Hamiltonian is diagonalized in the basis of 16 functions of the form
\begin{equation}
\label{sec2_1}
\psi ^{c/a}_{{\bf k}i\sigma}({\bf r})=\frac{1}{\sqrt{N_c}}\sum_{{\bf
R}_{c/a}} e^{i{\bf kR}_{c/a}} \varphi^{c/a}_{i}({\bf r}-{\bf
R}_{c/a})\, |\sigma \rangle ,
\end{equation}
where $\varphi^{c/a}_{i}({\bf r}-{\bf
R}_{c/a})$ with $i=p_x,p_y,p_z,s$ are the cation or anion atomic
orbitals centered on the lattice sites ${\bf R}_c$ or ${\bf R}_a$,
respectively, $N_c$ is the number of cation sites, and $|\sigma
\rangle$ with $\sigma = \pm \frac{1}{2}$ is the two dimensional spinor.
After diagonalizing the Hamiltonian matrix, for a given {\bf k} we
obtain the band energies $\varepsilon_{n{\bf k}}$ and corresponding
eigenfunctions
\begin{equation}
\label{sec2_2}
|q\rangle \equiv |n{\bf k}\rangle =\sum_{i}\sum_{\sigma}\sum_{p=c,a}
a^{p}_{{\bf k}i\sigma}
\psi ^{p}_{{\bf k}i\sigma}({\bf r}),
\end{equation}
where index $n=1,...16$ labels the band number.

The knowledge of the amplitudes $a^{p}_{{\bf k}i\sigma}$ enables us to
calculate the hybridization matrix elements $\langle d_i \sigma |h|
q \rangle$. We assume that there is only an overlap of 5$d$  RE orbitals
with six neighboring anions. The necessary values of inter atomic matrix
elements $\langle d_i|h|\varphi^{a}_{j} \rangle$ are calculated
according to Table 20-1 of Ref. \onlinecite{harrison} and can be
expressed using three constants $V_{sd\sigma}$, $V_{pd\sigma}$, and
$V_{pd\pi}$ defined as in Ref.~\onlinecite{harrison}:
\begin{eqnarray}
\label{sec2_3}
V_{ldm}=\eta_{ldm}\frac{\hbar^2r_d^{3/2}}{m_0d^{7/2}}
\end{eqnarray}
where $\eta_{sd\sigma}=-3.13$, $\eta_{pd\sigma}=-2.95$ and
$\eta_{pd\pi}=1.36$, $m_0$ is the bare electron's mass, $d=3.2$ \AA \
is the cation-anion distance in PbTe and $r_d$ is a fitting parameter
related to the radius of $5d$ RE orbital and it is of the order of 1
\AA.  In
calculations we put $r_d$=2.5 \AA \ and we obtain $V_{sd\sigma}$=-1.62
eV, $V_{pd\sigma}$=-1.51 eV and $V_{pd\pi}$=0.70 eV. These values are
close to the ones used in Ref.~\onlinecite{kacman1} in calculations of
EuTe band structure.\cite{kacman}

The summation over ${\bf k}_0$ in Eq.~(\ref{sec1_7}) is replaced by
the integration over  the Brillouin zone according to the formula:
\begin{equation}
\label{sec2_4}
\sum_{{\bf k}_0}\rightarrow 
V\int_{BZ}\frac{d^3{\bf k}}{\left(2\pi\right)^3}
\end{equation}
where $V$ is the volume of the crystal. \\ 
The value of $\epsilon_0$, the energy necessary to transfer an
electron from the top of the valence band to the $5d$ level of RE ion
is one of the most important parameters of the theory. According to
Ref.~\onlinecite{story2} the $5d$ level of Gd in PbTe lies about 0.2 eV
above the bottom of conduction band. Adding the value of the
energy gap in PbTe which is equal to 0.2 eV we obtain
$\epsilon_0$=0.4~eV for Gd in PbTe. From the resonant photo-emission
spectroscopy experiments\cite{kowalski} we know that $4f$ level of Eu
in PbTe lies near the top of the valence band while for Gd it is
placed about 10 eV below. That is why we expect that $\epsilon_0$ for Eu
is larger than for Gd. The precise value is not important because as
we will see the value of the coefficient $b_4$ decays very quickly
with increasing $\epsilon_0$ and if it is bigger than 1~eV the
contribution to $b_4$ from the present mechanism becomes negligible. \\
\begin{figure}
\begin{center}
\includegraphics[scale=0.3,angle=0]{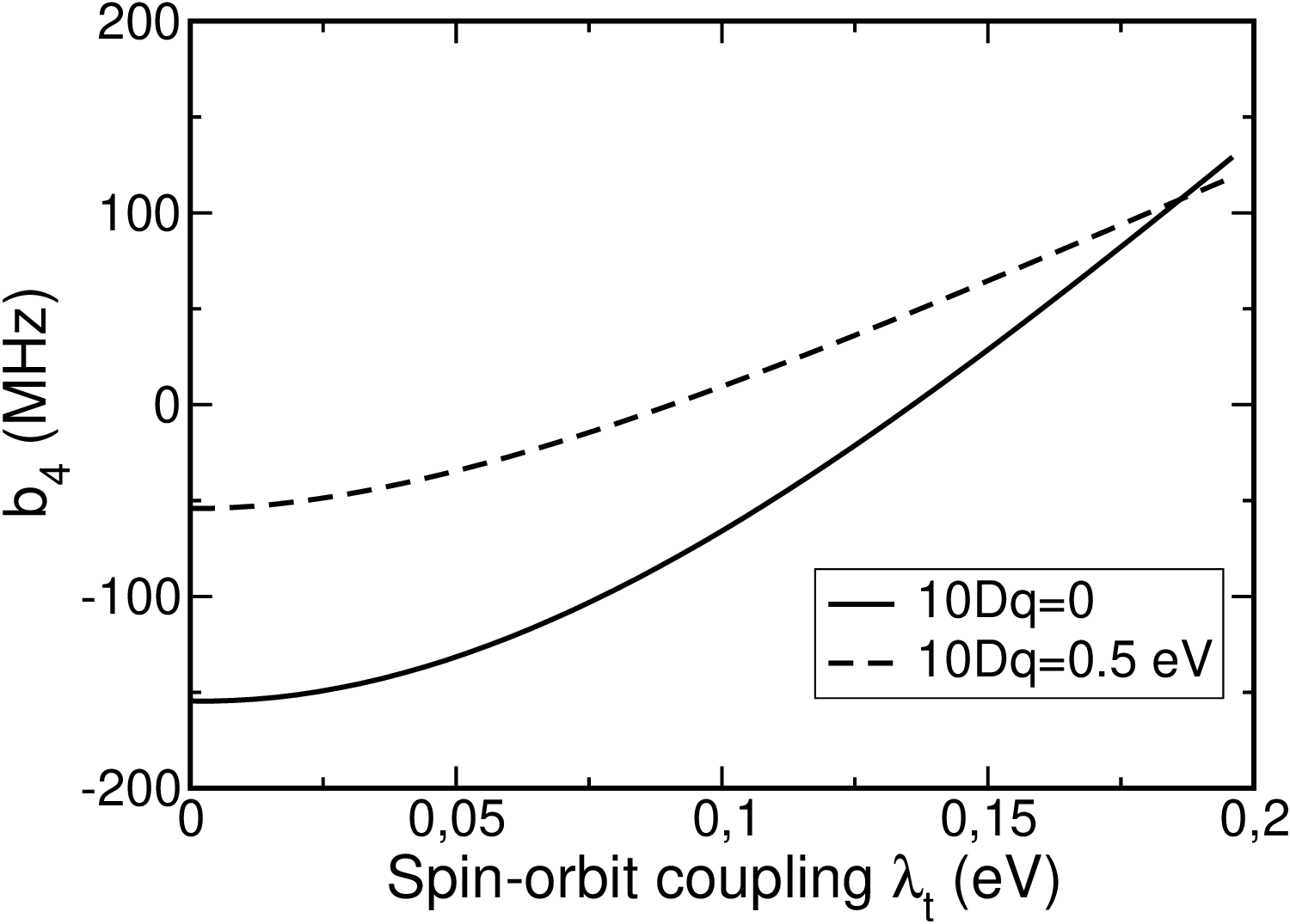}
\end{center}
\caption{\label{fig2} Dependence of $b_4$ on the spin - orbit constant
$\lambda_t$ for $\lambda_e$=0.1~eV, $J$=0.25~eV, $\epsilon_0$=0.35~eV
and for two different values of crystal field parameter $10Dq$. }
\end{figure}
In the literature we haven't found the values for the spin orbit
constants $\lambda_t$ and $\lambda_e$ for RE ions in PbTe. One may
find only value of single spin orbit constant $\lambda$. The existing
data are not very consistent, however. In Ref.~\onlinecite{methfessel} we
find $\lambda$=0.08~eV for Eu and $\lambda$=0.13~eV for Gd.
According to more recent theoretical relativistic calculations\cite{dolg}, $5d$ spin orbit
constant for free lanthanide ions is of the order of 0.06 eV. 
In our calculations we take $\lambda_e$=0.1 eV and in Fig.~\ref{fig2} we
present the results for $b_4$ as a function of $\lambda_t$ for
$\epsilon_0$=0.35~eV, $J$=0.25~eV and for two
different values of $10Dq=0$ and $10Dq$=0.5~eV. We clearly see that
the nonzero crystal field splitting of $5d$ strongly decreases $b_4$. Taking
into account the theoretical calculations\cite{francisco} of
$\lambda_t$ and $\lambda_e$  performed for $3d$ ions in crystals of
NaCl crystal structure, we expect that also in the present case the
real $\lambda_t$ should be less than $\lambda_e$.  \\
In Fig.~\ref{fig3} we show the decay of $b_4$ with increasing of
$\epsilon_0$ for $\lambda_t$=0.05~eV  and for two values of crystal
field parameter $Dq$.  \\
As it is clear from the above discussion the main problem in
calculations is the lack of knowledge of precise values of number of
necessary parameters of the model. However, it seems that the
estimations made above suggest that the proposed mechanism gives the
ground state splitting of the right order of magnitude and, moreover,
in the case of Gd in PbTe it gives proper sign of the coefficient
$b_4$. Also, from Fig.~\ref{fig3} we see that it is enough to
increase $\epsilon_0$ less than 1 eV to significantly decrease $b_4$,
which although negative, becomes very small and may be neglected. It
means that for such cases, for example for Eu in PbTe, the main
contributions to the splitting are due to other mechanisms, like
relativistic crystal field theory which generates positive sign of
$b_4$. \\
The above considerations indicate that the proposed model  should be
taken into account in theoretical analysis of EPR spectra. 
\begin{figure}
\vspace{1cm}
\begin{center}
\includegraphics*[scale=0.3,angle=0]{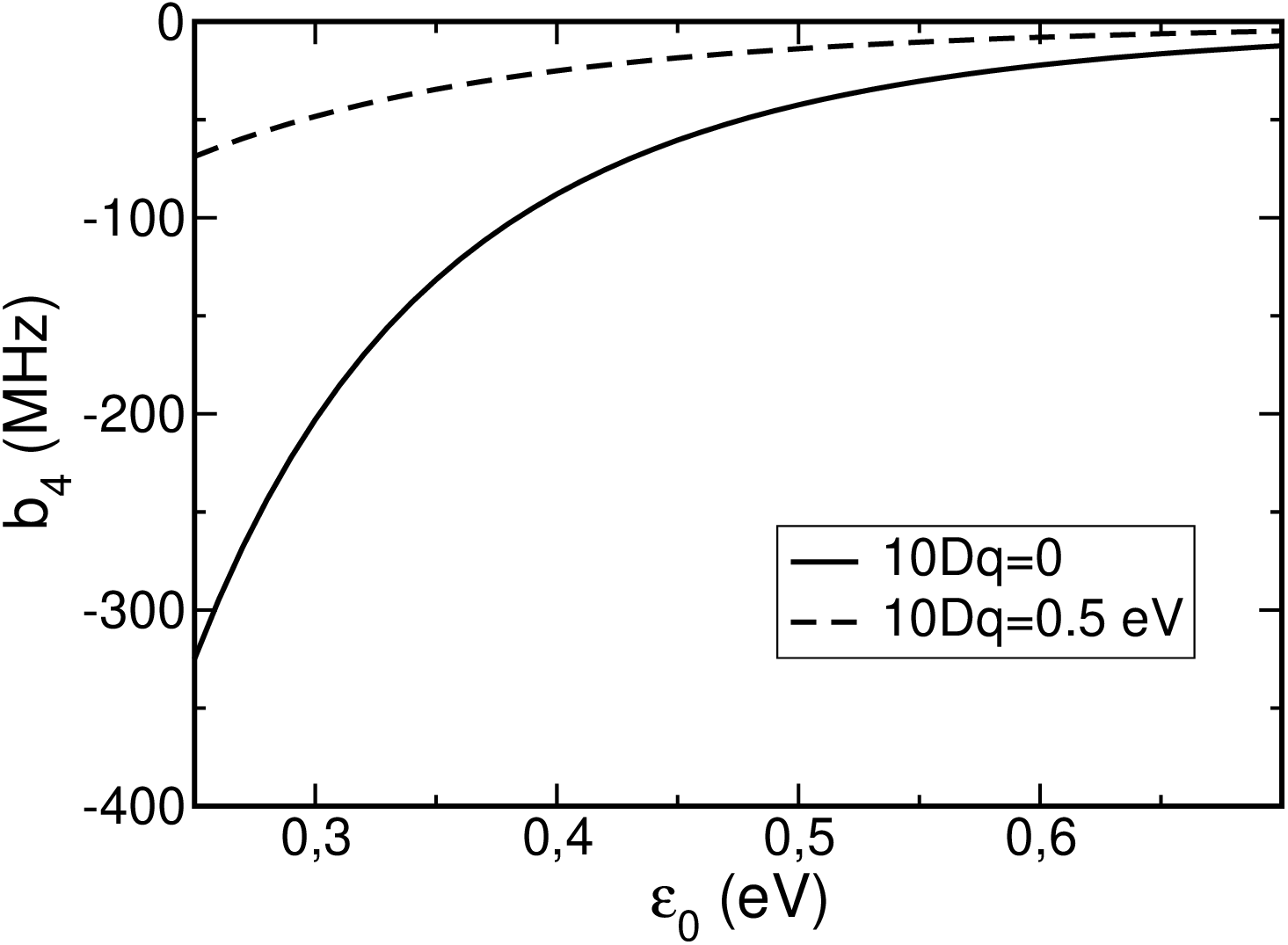}
\end{center}
\caption{\label{fig3} Dependence of $b_4$ on the transfer energy 
$\epsilon_0$ for $\lambda_e$=0.1~eV, $\lambda_e$=0.05~eV, $J$=0.25~eV 
and for two different values of crystal field parameter $10Dq$. }
\end{figure}

\section{Barnes, Baberschke and Hardiman model}
\begin{figure}
\vspace{1cm}
\begin{center}
\includegraphics*[scale=0.3,angle=0]{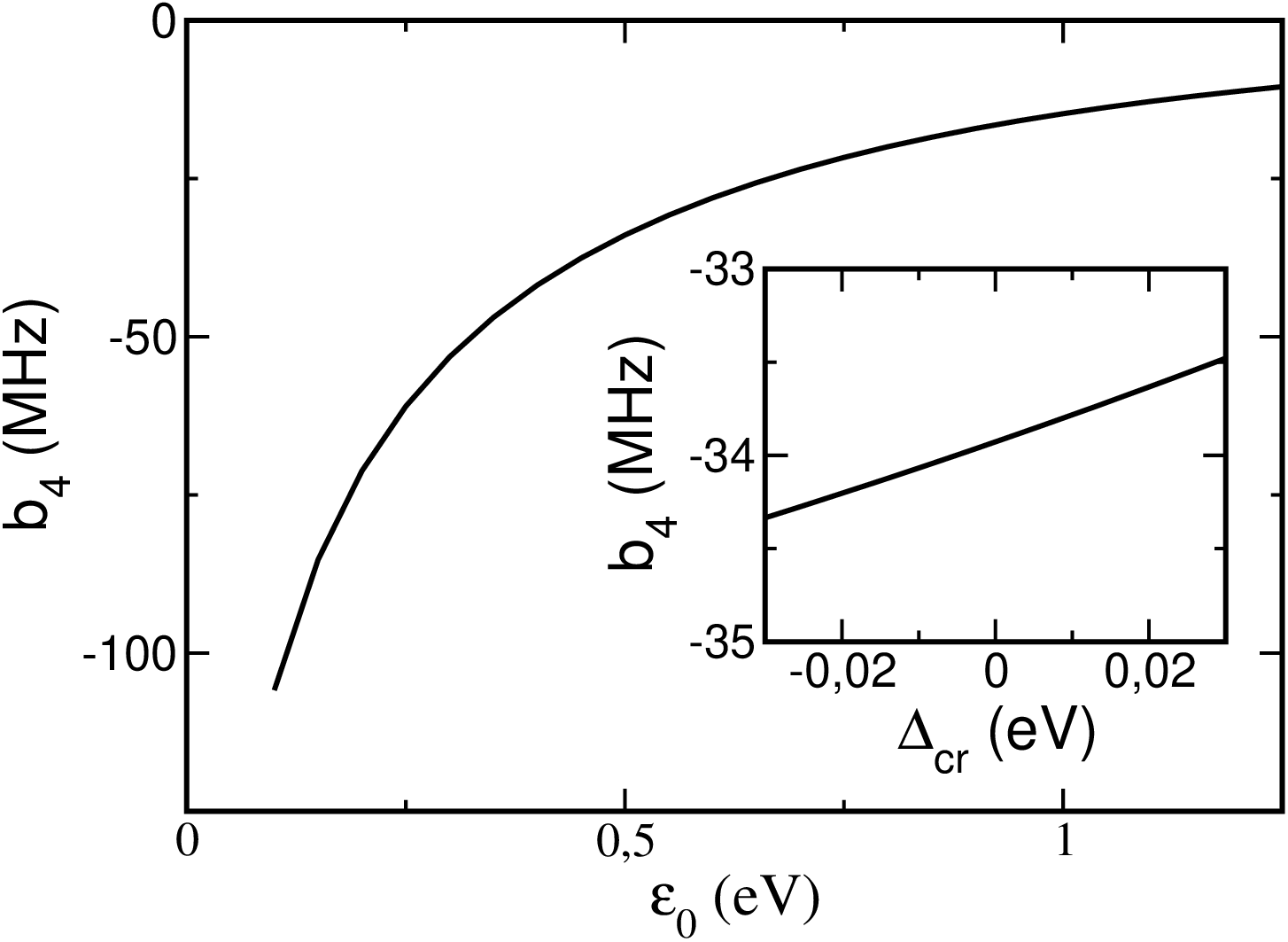}
\end{center}
\caption{\label{fig4} Dependence of $b_4$ on the transfer energy 
$\epsilon_0$ for $\lambda_{4f}$=0.0276~eV,  $V_{sd\sigma}$=-0.2 
eV, $V_{pd\sigma}$=-0.2 eV and $V_{pd\pi}$=0.1 eV and the crystal
field splitting $\Delta_{cr}$=0. 
The insert shows the dependence of $b_4$ on $\Delta_{cr}$ for
$\epsilon_0$=0.5 eV.  }
\end{figure}

As it has been discussed in the Introduction the main idea of the model
proposed by Barnes et. al.\cite{barnes} is to consider the
excited states of the system in which the number of electrons on the ion's $4f$ shell
changes by $\pm1$. Let us concentrate in this Section on processes
which lead to $4f^7 \leftrightarrow 4f^6$ transitions. The ground
state of the system, $|\ ^8S(J_z) \rangle$, is the same as in Section
II, namely the RE ion in $4f^7$  configuration plus the Fermi sea of
electrons. This eight fold degenerate state 
of the system is characterized by $-7/2 \le M\le 7/2 $ - the
projection of $4f^7$ spin 7/2 on a quantization axis which we take
along the (001) crystallographic direction. In the excited states we
have the ion in $4f^6$ configuration plus one additional electron
above the Fermi level characterized, like in Section II, by the quantum
number $q$. Assuming the validity of the Hund's rule for $4f^6$
configuration, L=3 and S=3, the Hamiltonian for the ion in excited
state reads:
\begin{equation}
\label{sec3_1}
H=\left(\lambda_{4}{\bf L}_{4}+\lambda_{5}{\bf L}_{5}+
\lambda_{25}{\bf L}_{25}+\lambda_{45}{\bf L}_{45}\right)\cdot{\bf S}+V_{cr}
\end{equation}
In the above equation the indices 2,4,5 correspond to the decomposition
of the $D^3$ representation into irreducible representation
$\Gamma_2$, $\Gamma_4$ and $\Gamma_5$ of the cubic
group.\cite{abragam} For example ${\bf L}_{4}$ is the angular momentum
operator between base functions of $\Gamma_4$ representation and ${\bf
L}_{45}$ between $\Gamma_4$ and $\Gamma_5$ base functions. ${\bf S}$
is the spin operator of the length S=3 and $V_{cr}$ is diagonal matrix
describing the crystal field potential.\\
Let us stress that from the physical point of view this is rather unreasonable to
describe spin - orbit interaction using four different constants
$\lambda_4$, $\lambda_5$, $\lambda_{25}$ 
and $\lambda_{45}$. The $4f$ shell is very localized and contrary to
$5d$ orbitals the influence of the neighbourhood should be negligible.
The only reason of introducing four constants instead of a
single one is, that  it is easier to understand
the influence of the crystal field on the coefficient $b_4$. \\
The important difference between our formulation of the model and that
of Barnes et. al.\cite{barnes} is in the form of hybridization
elements. We propose that
\begin{eqnarray}
\label{sec3_2}
\langle L_zS_z q |H|\ M \rangle= \\
(-1)^{L_z+1}\sum_{\sigma=\pm\frac{1}{2}}\sqrt{\frac{7/2+2\sigma M}{7}}\delta_{S_z,M-\sigma}
\langle q|h|\phi_{-L_z\sigma}\rangle. \nonumber
\end{eqnarray}
The state $|L_zS_z q\rangle$ is the excited state of the system in which the
projection on the quantization axis  of the total angular momentum
and spin of the ion are $L_z$ and $S_z$, respectively, and there is one
additional electron characterized by $q$ above the Fermi energy. The
element $\langle q|h|\phi_{-L_z\sigma}\rangle$ describes hybridization
between band state $q$ and the $4f$ spin orbital
$\phi_{-L_z\sigma}$. The coefficient $(-1)^{L_z+1}\sqrt{\frac{7/2+2\sigma
M}{7}}$, omitted in Ref.~\onlinecite{barnes}, may be derived using
explicit forms of antisymmetric many electron functions for ion's
states $|L_zS_z \rangle$ and $|\ M \rangle$\cite{comment}. \\
The further steps of calculation of the effective spin Hamiltonian
are, with minor modifications, very similar to those described in the
previous sections. Because of much higher complexity of the angular
momentum algebra for L=S=3 it is not easy to derive formulas analogous
to those from Section II. That is why the conclusions are based on
numerical calculations only, performed for PbTe crystal, the example 
semiconductor we use in the present paper. \\
In general, the results are similar to those for the model described
in Section II. First, we observe decrease of modulus of $b_4$ with the
increase of $\epsilon_0$, see Fig.~\ref{fig4}. In this Section
$\epsilon_0$ denotes the energy necessary to transfer an electron from
$4f$ shell to the Fermi energy level. Secondly, for 
$\epsilon_0$ and other 
parameters of the model kept constant, $b_4$ changes with the
crystal field potential parameter $\Delta_{cr}$, however
as we see in the insert of Fig.~\ref{fig4} these changes are not very
fast. $\Delta_{cr}$ is defined here as the total splitting of $4f^6$
manifold in the crystal field,
i. e. $\Delta_{cr}=E_{\Gamma_4}-E_{\Gamma_2}$.  The mechanism of
changes of $b_4$ with $\Delta_{cr}$ is similar to the one analysed in
Section II. From the performed numerical analysis it turns
out that $b_4\equiv$0 for $\lambda_{45}$=0. The relative energy
positions of $\Gamma_2$, $\Gamma_4$ and $\Gamma_5$ states, which change with
$\Delta_{cr}$ are important, because they decide about the
effectiveness of the transitions between them caused by spin orbit
interaction. However, comparing to $5d$, in the case of $4f$ shell
this influence is rather small. This is connected with completely
different ratio of spin-orbit to the crystal field strength. For the
$4f$ orbitals the effects caused by the crystal field are much smaller
than those due to spin orbit interaction. \\
This main difference between our results and those by Barnes et. al. is
that the authors of Ref.~\onlinecite{barnes}
overlooked the fact that even for zero crystal field potential we
obtain nonzero ground state splitting of the RE ion. This splitting, however,
may be achieved only if the
symmetry of the band wave functions and, consequently, the
proper symmetry of hybridization matrix elements is taken into
account. If these symmetries are neglected and the $4f$ - band
hybridization is described by a single constant $V_{fk}$, (see
Ref.~\onlinecite{barnes}) then, apart from the crystal
field potential, the rest of the  system's Hamiltonian  is invariant
with respect to operations of rotation group and  the ground state
remains degenerate. Then, indeed, in the first approximation, the
splitting will be proportional to 
the crystal field and this is the only contribution calculated in
Ref.~\onlinecite{barnes}. \\
On the basis of our analysis, see Fig.~\ref{fig4}, it is clear that in the case
of $4f$ - band states hybridization, the most important 
parameter of the model is not $\Delta_{cr}$ but $\epsilon_0$, the
energy necessary to transfer an electron from $4f$ shell to the
conduction band. 
\section{Conclusions}
In this paper we have proposed the model leading to the splitting of
the ground state of $^8S$ rare earth ions in crystals. The main
ingredients of the model are $4f-5d$ exchange intereaction, spin orbit
coupling for $5d$ electrons and their hybridization with the band
states. The numerical calculations have been performed for PbTe, a
semiconductor from IV-VI group of compounds. We have also limited
considerations to the $O_h$  symmetry of the ion's
neighbourhood. Of course, the model may be applied to any
semiconductor and, with slight modifications in hybridization matrix
elements, to lower symmetry cases. \\
In Section IV we have discussed the model proposed by Barnes
et. al. In particular, we have shown the importance of the careful
treatment of symmetry of the hybridization matrix. If the symmetry of
the problem of this kind ist lost due to too big simplifications in
the model's formulation we may also loose important, from the physical
point of view, class of solutions, as it has happened in
Ref.~\onlinecite{barnes}. \\
The general conclusion resulting from the analysis of both
models is inaccordance with the point of view presented in
Ref.~\onlinecite{barnes}, namely that the ground state splitting of
$^8S$ ions in crystals is 
not governed by their internal properties only, but the position of
the ion's $4f$ and $5d$ levels relative to the host crystal band
structure is at least of equal importance. \\
The model introduced in Section II, applied to Gd$^{3+}$ and Eu$^{2+}$ ions in PbTe
explains origin of opposite signs of the coefficients $b_4$ observed
in experiment.   

\begin{acknowledgments}
The author would like to thank Prof. R. R. Galazka and
Prof. T. Story  for helpful discussions and  for 
the critical reading of the manuscript. This work was partly supported by 
KBN Grant PBZ-044/P03/2001.  
\end{acknowledgments}


\begin{thebibliography}{22}
\bibitem[{com()}]{comment}
\bibinfo{note}{Strictly speaking Barnes et. al. in their considerations used
  only the matrix element for maximal spin, $\langle 33q|H|7/2\rangle$, thus
  this factor was unimportant, however, in more general treatment of the
  problem its presence is necessary.}
\end{thebibliography}

\end{document}